\documentclass[final,3p,times,twocolumn]{elsarticle}
\usepackage{amsmath}
\usepackage{graphicx}

\usepackage{amssymb}

\journal{Physica B}

\begin{document}

\begin{frontmatter}



\title{Ground-state Properties of Tonks-Girardeau Gas in One Dimensional Periodic Potential}



\author[label1]{Wang Zhao-liang}
\fntext[label1]{Corresponding author. Tel.:+86-551-3607061; fax:+86-551-3607061. E-mail address: wzlcxl@mail.ustc.edu.cn}
\author{Wang An-Min}
\author{Li Xue-Chao}
\address{Department of Modern Physics, University of Science and
Technology of China, Hefei, Anhui 230026, China}

\begin{abstract}
The relations among the occupation number of the lowest natural orbital (ONLNO), momentum distributions (MD) and off-diagonal long-range element (ODLRE) of the reduced single-particle density matrix (RSPDM) are studied while Tonks-Girardeau gas in one dimensional periodic potential is in the ground state. For $N$-body systems of large enough, RSPDM and its lowest natural orbital do not vary with $N$ in overlapped areas in commensurate and incommensurate cases correspondingly. In commensurate case, the ODLRE is exponential attenuation with $N$, which results in that the ONLNO and MD are invariant with $N$. While in contrast, in incommensurate case, the off-diagonal elements are inversely proportional to $\sqrt{N}$, which results in the different behavior of the ONLNO and MD.


\end{abstract}

\end{frontmatter}



\section{Introduction}\label{sec:level1}

With the development of realizing quasi-one dimensional cold atomic systems in experiment\cite{BP,TK}, one dimensional systems which could be exactly solved have again drawn people's attention. One of these is Tonks-Girardeau\cite{Tonks,Girardeau1960} (TG) gas which consists of bosons with ``impenetrable" core repulsive interactions. Actually, at low temperatures and densities, Bose system will act as a TG system\cite{Olshanii,Petrov,Dunjko}. TG gas is first a Bose gas and to some extent it also displays some properties just like a Fermi gas\cite{Pezer,Wang}.

As the TG gas model can be exactly solved via Fermi-Bose mapping\cite{Girardeau1960,Girardeau1965,Girardeau2000} which does not depend on the external potential, many models with different external potentials have been well studied theoretically. $\delta$-split harmonic potential model was studied by J. Goold and Th. Busch\cite{Goold}. They showed that the scaling of the occupation number depends on whether one has an even or odd number of particles. Kronig-Penney potential model was investigated by Lin, etc.\cite{Lin} and Wei, etc.\cite{Wei}, who pointed that there are two different phases in periodic potentials. One is Mott insulator phase that the ratio of the number of bosons to wells $N/M$ is an integer, the other is boson conductor phase which $N/M$ is a fractional number. The two phases displayed entirely different physical properties\cite{Wei}. Cosine potential model was considered by Z. L. Wang and A. M. Wang\cite{Wang}. There are two advantages of choosing cosine potential. First, it is easier to prepare ground state in Mott insulator phase. Second, it is easier to prepare cosine potential than $\delta$ potential and it is convenient to control amplitude $B$ and frequency $\omega$. In this work, we continue studying cosine potential.

Many properties of the TG gas and its corresponding Fermi gas are always identical\cite{Girardeau1960,Wang,Girardeau2000}, such as the average value of particle coordinates, potential energies, system total momenta, single-particle densities and pair distribution functions. But other properties are quite different, such as RSPDM and MD functions $n(k)$\cite{Pezer, Wang, Lin, Wei, Lenard, Girardeau2001, Cazalilla, Berman, Rigol}. Before R. Pezer and H. Buljan\cite{Pezer} introduced a method for calculating the RSPDM of a TG gas, it is difficult to calculate the RSPDM and the corresponding MD for large number systems\cite{Lenard,Girardeau2001,Cazalilla,Berman,Rigol}. In this manuscript, we mainly use Pezer's method\cite{Pezer} to study non-local properties of the TG gas.

Generally speaking, there are three discrimination methods in judging whether a boson system occurs Bose-Einstein Condensate (BEC), such as the exhibition of off-diagonal long-range order (ODLRO), the macroscopic occupation numbers of the lowest natural orbital and the zero-momentum state\cite{Girardeau,Penrose,Yang}. These physical quantities are all derived from RSPDM, so they are not independent and we believe that they have some relations among them. Through the studies of properties of TG gas in periodic potential, we have found the relations and deduced one form in the text.

This work is organized as follows. In Sec.2, we introduce the model Hamiltonian and its computation processes. In Sec.3, we study the ground state properties of the TG gas by numerical calculation and theoretical explanation. In Sec.4, brief conclusions are given.

\section{Model Hamiltonian and wave functions}\label{sec:level2}

In a TG system, the boson is assumed to have an ``impenetrable'' hard core
characterized by a radius of $a$. From Girardeau's work, with the
hard core radius $a\rightarrow0$, the interparticle interaction is
given by
\begin{equation}
U(x_{_i},x_{_j})=\begin{cases}
0, & \text{$x_{_i} \neq x_{_j}$},\\
\infty, &\text{$x_{_i}=x_{_j}$}.
\end{cases}
\end{equation}

Such an interparticle interaction could be represented by the
following subsidiary condition on the wave function $\psi$ :
\begin{equation}\label{eqn2}
\psi(x_{_1},\cdots,x_{_N},t)=0\,\,\,\mathrm{if}\,\,\,\hspace{\stretch{1}}x_{_i}=x_{_j},\hspace{\stretch{0.3}}1\leqslant
i<j\leqslant N.
\end{equation}

With Girardeau's Bose-Fermi mapping\cite{Girardeau1960}, the wave function of TG system is given by wave function of noninteracting spinless Fermions multiply an antisymmetric factor.
\begin{equation}\label{eqn7}
\psi(x_{_1},\cdots,x_{_N},t)=A(x_{_1},\cdots,x_{_N})\psi^F(x_{_1},\cdots,x_{_N},t),
\end{equation}
in which
\begin{equation}
A(x_1,\cdots,x_N)\equiv \prod^N_{i>j}\mathrm{sgn}(x_i-x_j),
\end{equation}
\begin{equation*}
\mathrm{sgn}(x)\equiv \frac{x}{|x|}=\begin{cases} 1, & \text{$x>0$},\\ -1,
&\text{$x<0$}.
\end{cases}
\end{equation*}

Now we study the case that magnetized bosons in an external magnetic field $B(x)=-B\cos(2\omega x)$. Without considering the interparticle interactions,  the total Hamiltonian is written as
\begin{equation}\label{eqn1}
\hat{H}=\sum^N_{i=1} \left[-\frac{\hbar^2}{2m}\frac{\partial^2}{\partial
x^2_i}+V(x_i) \right].
\end{equation}
in which
\begin{equation}\label{eqn2}
V(x)=\mu B\cos(2\omega x),\qquad -\frac{L}{2}\leqslant x\leqslant \frac{L}{2}
\end{equation}

We suppose $L=M\pi/\omega$, where $M$ is an integer, $\mu$ is the magnetic moment of atom and $\omega$ is the circular frequency. For the sake of simplicity, we assume that both $N$ and $M$ are odd (As a result of the introduction of $A(x_{_1},\cdots,x_{_N})$, the wavefunction $\psi$ of TG system is periodic if $N$ is odd while otherwise antiperiodic\cite{Girardeau1960,Yukalov}). The single-particle Schrodinger equation in $x\in[-\frac{L}{2},\frac{L}{2}]$ is written as
\begin{equation}\label{eqn3}
\left[-\frac{\hbar^2}{2m}\frac{\partial^2}{\partial x^2}+\mu
B\cos(2\omega x)\right]\varphi_m(x)=E_{\alpha}\varphi_m(x).
\end{equation}
Substitute $z=\omega x$, $q=\dfrac{m\mu B}{\hbar^2\omega^2}$ and $\lambda=\dfrac{2mE}{\hbar^2\omega^2}$ into equation (\ref{eqn3}), then it becomes
\begin{equation}\label{eqn4}
\frac{d^2\varphi(z)}{d z^2}+[\lambda-2q\cos(2z)]\varphi_m(z)=0,
\end{equation}
\begin{equation*}
z\in\left[-\frac{\omega L}{2},\frac{\omega L}{2}\right].
\end{equation*}

Using Bloch's theorem and periodic boundary conditions $\varphi_m(z)=\varphi_m(z+\omega L)$, $\varphi_m(z)$ is periodic with period $\pi$, so we can expand it in Fourier series:
\begin{equation}\label{eqn5}
\varphi_m(z)=\exp(i\nu z)\sum_nc_{_n}\exp(i2nz),
\end{equation}
\begin{equation*}
\nu=\frac{2l}{M},\;\;\left(l\in0,\pm1,\cdots,\pm\frac{M-1}{2}\right).
\end{equation*}

Then substitute (\ref{eqn5}) into (\ref{eqn4}), we obtain an infinite symmetric tridiagonal matrix equation which is the eigen equation.

\begin{multline}\label{eqn6}
\negthickspace\negthickspace\begin{pmatrix}
\ddots & \cdots &\negthickspace \cdots \negthickspace & \cdots \negthickspace & \cdots\negthickspace & \cdots & \cdots \\
\cdots & (\nu-4)^2 &\negthickspace q \negthickspace & 0 \negthickspace & 0\negthickspace & 0 & \cdots\\
\cdots & q &\negthickspace (\nu-2)^2 &\negthickspace q &\negthickspace 0\negthickspace & 0 & \cdots \\
\cdots & 0 &\negthickspace q &\negthickspace \nu^2 &\negthickspace q\negthickspace & 0 & \cdots\\
\cdots & 0 &\negthickspace 0\negthickspace & q &\negthickspace (\nu+2)^2\negthickspace & q & \cdots\\
\cdots & 0 &\negthickspace 0 &\negthickspace 0 &\negthickspace q &\negthickspace (\nu+4)^2 & \cdots\\
\cdots & \cdots &\negthickspace \cdots &\negthickspace \cdots &\negthickspace \cdots\negthickspace & \cdots & \ddots
\end{pmatrix}\\
\times\begin{pmatrix} \vdots \\ c_{_{-2}} \\ c_{_{-1}} \\ c_{_0} \\
c_{_1} \\ c_{_2} \\\vdots
\end{pmatrix}
=\lambda
\begin{pmatrix}
\vdots \\ c_{_{-2}} \\ c_{_{-1}} \\ c_{_0} \\ c_{_1} \\ c_{_2}
\\\vdots
\end{pmatrix}.
\end{multline}

By truncating the matrix in each direction (centered at the smallest diagonal element $\nu^2$) at sufficiently large dimensions, approximations to the desired eigenvalues and eigenfunctions can be obtained to any desired precision. For more detailed computation process, please refer to the references\cite{Wang,Randall,Blanch,Dingle}.

The wave functions of the TG system are given by the Slater determinant
\begin{equation}\label{eqn7.1}
\psi=A(x_{_1},\cdots,x_{_N})\frac{1}{\sqrt{N!}}\mathrm{Det}_{m,j=1}^N\left[\varphi_m\left(x_j\right)\right].
\end{equation}

When the temperature is zero, the system will be in the ground state --- $N$ particles in the $N$ lowest eigenstates, respectively. In this work, we mainly consider the ground state, so we should choose the lowest $N$ single-particle eigenstates in Eq. (\ref{eqn7.1}).

\section{Relations Among Some Properties}\label{sec3}

It is easy to prove that single particle density and pair distribution functions are always the same for TG gas and its mapping spinless Fermi gas\cite{Girardeau1960,Wang,Lin,Yukalov}. Here we just consider properties that differ from the two systems, such as RSPDM, MD, natural orbital and its occupation numbers.

\subsection{Reduced Single-particle Density Matrix}\label{subsec:level1}

The reduced single-particle density matrix, with normalization $\int\rho(x,x)\, \mathrm{d}x=N$, is given by
\begin{equation}\label{eqn8}
\rho(x,x')\negthickspace\equiv\negthickspace N\negthickspace\int\negthickspace\psi^*(x,x_{_2},\negthickspace\cdots\negthickspace,x_{_N})\psi(x',x_{_2},\negthickspace\cdots\negthickspace,x_{_N})
dx_{_2}\negthickspace\cdots\negmedspace dx_{_N}.
\end{equation}

\begin{figure}[!h]
\center
\includegraphics[scale=0.35]{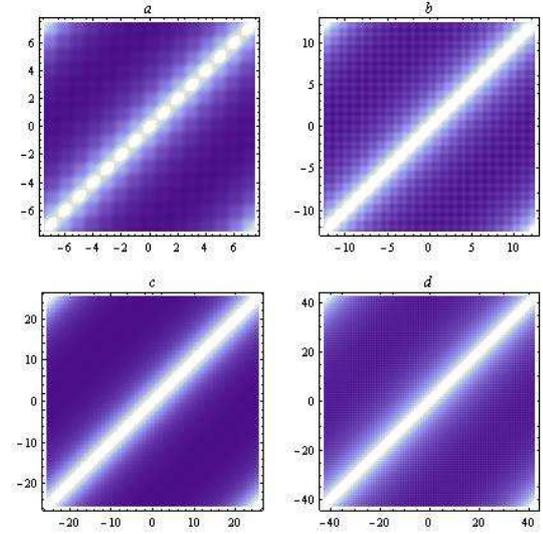}
\caption{(Color online) RSPDM $\rho(x,x')$ of the TG gas in external periodic magnetic field for different particles in commensurate and incommensurate cases, $q=0.1,\;\omega=1$. The abscissa axis $x$ and the ordinate axis $x'$ are in units of $\pi$. (a) $N=15,N/M=1$; (b) $N=15,N/M=3/5$; (c) $N=51,N/M=1$; (d) $N=51,N/M=3/5$.}.\label{fig1}
\end{figure}
\begin{figure}[!h]
\center
\includegraphics[scale=0.28]{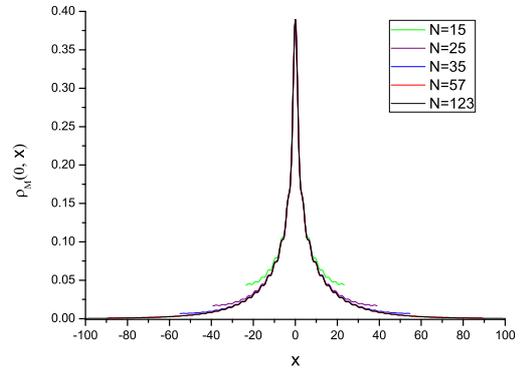}
\caption{(Color online) Off-diagonal elements $\rho_{_M}(0,x)$ in commensurate case, $N/M=1$. $q=0.1,\;\omega=1$. From the higher to the lower: $N=15,25,35,57,123$.}
\label{fig2}
\end{figure}
\begin{figure}[!h]
\center
\includegraphics[scale=0.28]{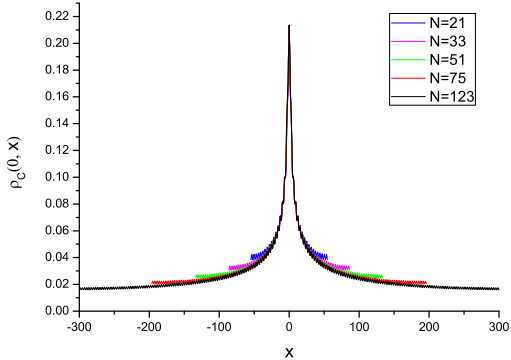}
\caption{(Color online) Off-diagonal elements $\rho_{_C}(0,x)$ in incommensurate case, $N/M=3/5$. $q=0.1,\;\omega=1$. From the higher to the lower: $N=21,33,51,75,123$.}
\label{fig3}
\end{figure}

The RSPDM, which expresses self-correlation, reflects the probability that having found a particle at position $x$ and at the same time detecting the particle at position $x'$. Many important observable physical quantities are defined by the RSPDM, such as the occupation numbers of natural orbital and the MD, but it is hard to calculate $\rho(x,x')$ by the definition of equation (\ref{eqn8}), even if numerical calculate. Fortunately, it can be expressed in terms of the dynamically evolving single-particle basis raised by R. Pezer and H. Buljan\cite{Pezer}:
\begin{equation}\label{eqn9}
\rho(x,x')=\sum_{i,j=1}^N\varphi_i^*(x)A_{ij}(x,x')\varphi_j(x').
\end{equation}
The $N\times N$ matrix $\mathbf{A}(x,x')=\{A_{ij}(x,x')\}$ is
\begin{equation}\label{eqn10}
\mathbf{A}(x,x')=(\mathbf{P}^{-1})^T\mathrm{Det}\mathbf{P},
\end{equation}
where the entries of the matrix $\mathbf{P}$ are $P_{ij}(x,x')=\delta_{ij}-2\int_x^{x'}dy\,\varphi_i^*(y)\varphi_j(y)$, and we have assumed $x<x'$ without loss of generality. In addition, the RSPDM in periodic potential satisfies \cite{Wei}
\begin{equation}\label{eqn11}
\rho(x+T,x'+T)=\rho(x,x'),
\end{equation}
where $T$ is the period of external periodic potential, and in this work $T=\pi/\omega$. With this formula, we can calculate considerable large systems of the TG gas.

Figure \ref{fig1} displays contour plot of RSPDM for both the Mott insulator and boson conductor phase. For both the two phases in large systems, $\rho(x,x')$ does not vary with $N$ in overlapped areas for fixed $N/M$. In order to illustrate this view, we have plotted $\rho_{_M}(0,x)$ for $N=M$ in figure \ref{fig2} and $\rho_{_C}(0,x)$ for $N/M=3/5$ in figure \ref{fig3} (The reason of choosing $\rho(0,x)$ is to avoid the boundary effect\cite{Wei,Wang}, which results from the periodic boundary conditions), where the subscripts $M$ and $C$ stand for Mott insulator phase and boson conductor phase respectively. As the system is in Mott-insulator phase, all particles tend to be localized and have little correlations in long distance. Therefore, $\rho(0,x)$ decreases quickly in Mott-insulator phase while relatively smooth in boson conductor phase.

\subsection{Off-diagonal Elements}\label{subsec:level2}

ODLRO was suggested by Yang\cite{Yang} as the type of ordering for superfluids, such as BEC and electron pairs in superconductors.
In the thermodynamic limit and $|x-x'|\rightarrow\infty$, if the off-diagonal element $\rho(x,x')=0$, then ODLRO is not present and there is no BEC; Otherwise, ODLRO is present and there is BEC\cite{Girardeau,Penrose,Yang}.
\begin{figure}[!h]
\center
\includegraphics[scale=0.28]{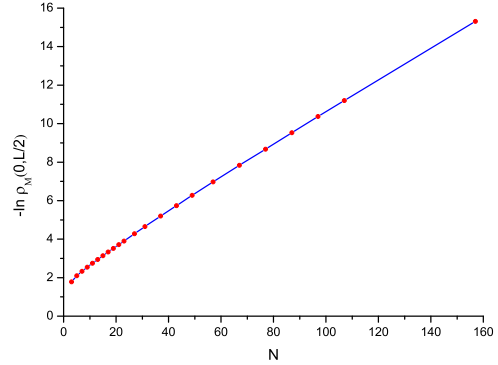}
\caption{(Color online) Off-diagonal elements $-\ln \rho_{_M}(0,L/2)$ as a function of $N$ in commensurate case, $N=M$, $q=0.1,\;\omega=1$.}
\label{fig4}
\end{figure}
\begin{figure}[!h]
\center
\includegraphics[scale=0.28]{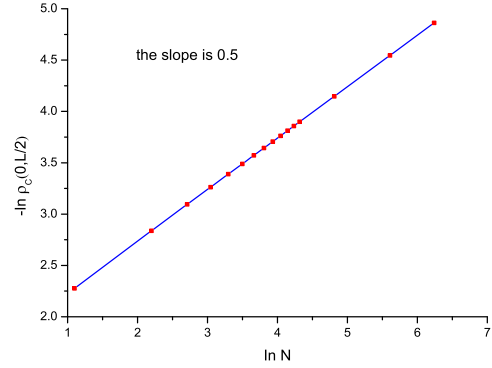}
\caption{(Color online) Off-diagonal elements $-\ln \rho_{_C}(0,L/2)$ as a function of $\ln N$ in incommensurate case, $N/M=3/5$, $q=0.1,\;\omega=1$.}
\label{fig5}
\end{figure}

Considering the periodic boundary conditions, we assume $\rho(0,L/2)$ to be a typical value of off-diagonal elements in the thermodynamic limit. The off-diagonal elements $\rho_{_M}(0,L/2)$ and $\rho_{_C}(0,L/2)$ as a function of particle number $N$ are shown in figure \ref{fig4} and \ref{fig5} respectively. Based on these numerical data, empirical correlation on ODLRE was obtained via regression. The results are as follows,
\begin{subequations}\label{eqn12}
\begin{equation}
\rho_{_M}(0,L/2)=C_1\exp(-\kappa_1 N),
\end{equation}
\begin{equation}
\rho_{_C}(0,L/2)=C_2N^{-1/2},\qquad
\end{equation}
\end{subequations}
where $C_1$, $C_2$ and $\kappa_1$ are constants and they vary with $q$. From equation (\ref{eqn12}), we know the off-diagonal elements $\rho(0,L/2)$ are zero in both the commensurate and incommensurate cases in the thermodynamic limit. Therefore, there are no ODLRO and BEC in both commensurate and incommensurate cases of TG gas in the external periodic potential.

\subsection{Occupation Numbers and Natural Orbital}\label{subsec:level3}

\begin{figure}[!h]
\center
\includegraphics[scale=0.28]{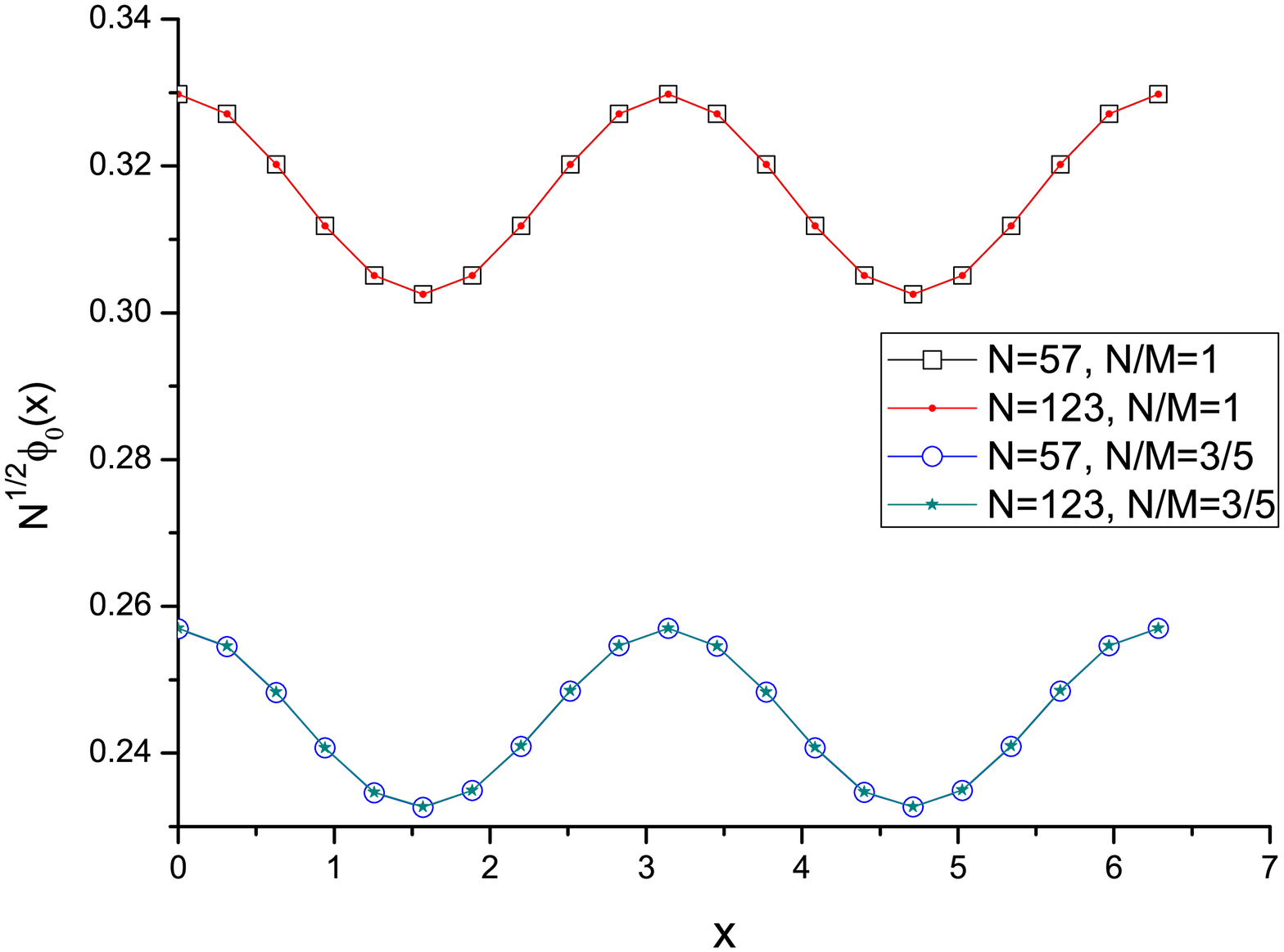}
\caption{(Color online) Scaled lowest natural orbital $\phi_0(x)$ for $N=57$ and $N=123$ in commensurate case $N=M$ and incommensurate case $N/M=3/5$, $q=0.1,\;\omega=1$.}
\label{fig6}
\end{figure}

The occupation numbers and the natural orbital are defined as
\begin{equation}\label{eqn13}
\int dx'\rho(x,x')\phi_i(x')=\lambda_i\phi_i(x),
\end{equation}
where $\lambda_i$ represents the occupation number of the natural orbital $\phi_i$ and $\sum_i\lambda_i=N$. For simplicity, we label the eigenvalues $\lambda_i$ in a descending order: $\lambda_0>\lambda_1>\lambda_2>\cdots$. The corresponding state $\phi_0$ of $\lambda_0$ is the condensate state of bosons. For the ground state, $\phi_0$ is the lowest natural orbital and the corresponding occupation number $\lambda_0$ is the largest. For large systems, from equation (\ref{eqn12}) we know the off-diagonal elements with long distances have few contributions to $\phi_0$, so the scaled lowest natural orbital $\sqrt{N}\phi_0(x)$ are the same in overlapped areas for different $N$ in commensurate and incommensurate cases separately. The lowest natural orbital is periodic with period $T=\pi/\omega$, which is the same as the external periodic potential. We have plotted two periods in figure \ref{fig6}.

In a macroscopic system, the presence or absence of BEC is determined by the behavior of $\rho(x,x')$ as $|x-x'|\rightarrow\infty$ or the largest eigenvalue $\lambda_0$ of $\rho(x,x')$ proportional\cite{Girardeau,Penrose,Yang} to $N$. Both the occupation number $\lambda_0$ and the off-diagonal element $\rho(0,L/2)$ are determined by $\rho(x,x')$, hence they must have some deep relations. For large systems, it is convenient to use $\phi_0(N,x)$ and $\lambda_0(N)$ to represent the lowest natural orbital and its occupation number of $N$-body system. Then,
\begin{equation}\label{eqn14}
   \begin{split}
   \lambda_0(N)=&\frac{1}{\phi_0(N,x)}\int_{-\frac{L}{2}}^{\frac{L}{2}}dx'\rho(x,x')\phi_0(N,x')\\
   =&\frac{1}{\phi_0(N,x)}\int_{-(1-\frac{1}{N})\frac{L}{2}}^{(1-\frac{1}{N})\frac{L}{2}}dx'\rho(x,x')\phi_0(N,x')\\
   +&\frac{1}{\phi_0(N,x)}\int_{(1-\frac{1}{N})\frac{L}{2}}^{(1+\frac{1}{N})\frac{L}{2}}dx'\rho(x,x')\phi_0(N,x').
   \end{split}
\end{equation}
From the definition of $\lambda_i$, we know the first term of the right hand in equation (\ref{eqn14}) is just $\lambda_0(N-1)$, where we have used the invariance of $\rho(x,x')$ and the accompanying $\sqrt{N}\phi_0(N,x)$ for fixed $N/M$. The value of $\lambda_0(N)$ has nothing to do with $x$, so the result of the second term in equation (\ref{eqn14}) should not include $x$. We set $x=0$ without loss of generality. Using the mean value theorem for integrals, the second term becomes
\begin{equation}
   \frac{1}{\phi_0(N,0)}\int_{(1-\frac{1}{N})\frac{L}{2}}^{(1+\frac{1}{N})\frac{L}{2}}dx'\rho(0,x')\phi_0(N,x')\approx C\rho_0(0,L/2).
\end{equation}
where $C$ is a constant. Rewriting equation (\ref{eqn14}), we obtain the relation between the occupation number of the natural orbital and the off-diagonal elements,
\begin{equation}\label{eqn15}
\frac{\Delta\lambda_0}{\Delta N}=\lambda_0(N)-\lambda_0(N-1)=C\rho(0,L/2).
\end{equation}
In the thermodynamic limit and regard $N$ as continuous, equation (\ref{eqn15}) becomes
\begin{subequations}\label{eqn16}
\begin{equation}\label{eqn16a}
\frac{d\lambda_0}{dN}=C\rho(0,\infty),
\end{equation}
or
\begin{equation}
\frac{d\lambda_0}{dN}=C\lim_{|x-x'|\rightarrow\infty}\mathrm{lim\enskip therm}\enskip \rho(x,x'),
\end{equation}
\end{subequations}
where ``lim therm" means ``thermodynamic limit''. From equation (\ref{eqn16}) we know that the behavior of the largest eigenvalue $\lambda_0$ and the off-diagonal elements $\rho(x,x')$ as $|x-x'|\rightarrow\infty$ are one-to-one correspondence.
\begin{figure}[!h]
\center
\includegraphics[scale=0.28]{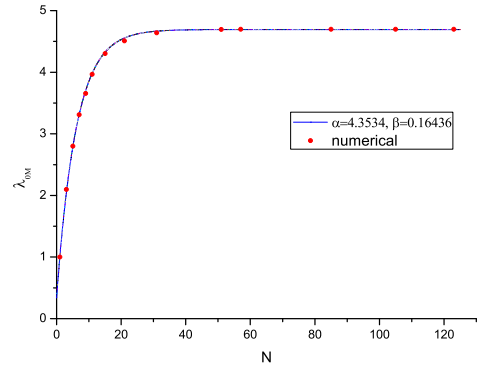}
\caption{(Color online) ONLNO as a function of $N$ in commensurate case, $q=0.1,\;\omega=1$.}
\label{fig7}
\end{figure}
\begin{figure}[!h]
\center
\includegraphics[scale=0.28]{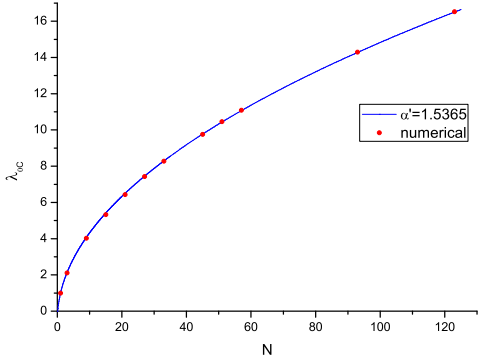}
\caption{(Color online) ONLNO as a function of $N$ in incommensurate case $N/M=3/5$, $q=0.1,\;\omega=1$.}
\label{fig8}
\end{figure}

Substitute equation (\ref{eqn12}) into (\ref{eqn16a}), we can get the expressions of $\lambda_{0M}$ and $\lambda_{0C}$ in commensurate and incommensurate cases separately,
\begin{subequations}\label{eqn17}
\begin{equation}
\lambda_{0M}=\alpha[1-\exp(-\beta N)]+\gamma,
\end{equation}
\begin{equation}\label{eqn17.2}
\lambda_{0C}=\alpha'\sqrt{N}+\gamma',
\end{equation}
\end{subequations}

where $\alpha$, $\beta$, $\gamma$, $\alpha'$ and $\gamma'$ are constants. $\gamma$ and $\gamma'$ are the correction terms resulted from the periodic boundary conditions. In both the two phases, $\lambda_{0M}$ or $\lambda_{0C}$ equals to 1 while $N=1$, so we can choose $\gamma=1-\alpha[1-\exp(-\beta)]$ and $\gamma'=1-\alpha'$. We have plotted $\lambda_{0M}$ and $\lambda_{0C}$ and their fitted curves with equation (\ref{eqn17}) in figure \ref{fig7} and \ref{fig8}. They clearly show that the numerical result is consistent with our theoretical equations.

Generally speaking, $\alpha'>1$ in one dimensional periodic TG systems. In this case, we can prove that equation (\ref{eqn17.2}) is equivalent to Girardeau's result\cite{Girardeau2001} $\lambda_{0C}\sim N^\sigma$, where $\sigma$ is a constant that slightly larger than $0.5$ and it gradually diminishes to $0.5$ in the thermodynamic limit\cite{Wei}.

\subsection{Momentum Distribution}\label{subsec:level3}

\begin{figure}[!h]
\center
\includegraphics[scale=0.25]{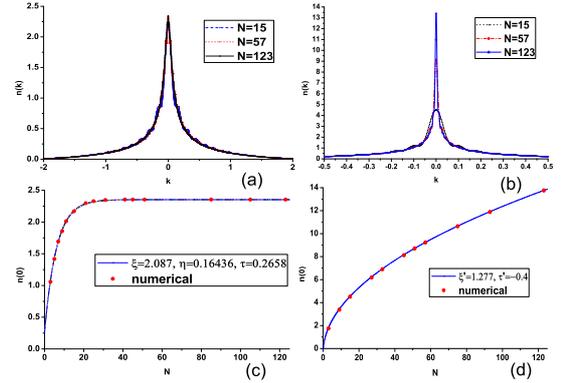}
\caption{(Color online) Normalized MD, $q=0.1,\;\omega=1$. (a) and (b): Normalized MD for different $N$ in commensurate and incommensurate ($N/M=3/5$) cases. (c) and (d): $n(0)$ as a function of $N$ in commensurate and incommensurate ($N/M=3/5$) cases.}
\label{fig9}
\end{figure}

The normalized momentum distribution, related to the RSPDM, is defined as
\begin{equation}\label{eqn18}
n(k)=\frac{1}{2\pi N}\int\rho(x,x')e^{-ik(x-x')}\,dx\,dx',
\end{equation}
with the normalization
\begin{equation*}
\int n(k)\,dk=1.
\end{equation*}
Actually, MD is the Fourier transformation of the RSPDM and they are one-to-one correspondence. Therefore, the behavior of MD will be determined by the behavior of the elements of RSPDM. We have plotted $n(k)$ in commensurate and incommensurate cases in figure \ref{fig9} (a) and (b) separately, in which the platforms will diminish as $N$ increases and they will vanish in the thermodynamic limit.

The quantity $n(k)$ is the probability density of particles, on the average, in the momentum $k$ state. If a momentum state $k_0$ has a macroscopic occupation such that $n(k_0)\propto N$, then the system will exhibit BEC\cite{Saarela}. In this work, it is clear to see that $k_0=0$ is the state that has the largest occupation from figure \ref{fig9} (a) and (b). With the method used in equation (\ref{eqn14}), we can prove that $n(0)$ has the similar relationships as equation (\ref{eqn16}),
\begin{subequations}\label{eqn19}
\begin{equation}\label{eqn19a}
\frac{dn(0)}{dN}=C'\rho(0,\infty),
\end{equation}
or
\begin{equation}
\frac{dn(0)}{dN}=C'\lim_{|x-x'|\rightarrow\infty}\mathrm{lim\enskip therm}\enskip \rho(x,x'),
\end{equation}
\end{subequations}
where $C'$ is an constant.

Substitute equation (\ref{eqn12}) into (\ref{eqn19a}), $n_{_M}(0)$ and $n_{_C}(0)$ in the commensurate and incommensurate cases can be expressed as
\begin{subequations}\label{eqn20}
\begin{equation}
n_{_M}(0)=\xi[1-\exp(-\eta N)]+\tau,
\end{equation}
\begin{equation}
n_{_C}(0)=\xi'\sqrt{N}+\tau',
\end{equation}
\end{subequations}
where $\xi$, $\xi'$, $\eta$, $\tau$ and $\tau'$ are constants. $\eta$ equals to $\beta$ while the external periodic potential and particles are the same, where $\beta$ is a parameter in equation (\ref{eqn17}). $\tau$ and $\tau'$ are the correction terms resulted from the periodic boundary conditions. $n_{_M}(0)$, $n_{_C}(0)$ and their fitted curves with (\ref{eqn20}) are shown in figure \ref{fig9} (c) and (d). It is easily to see that the number of particles occupying zero-momentum state $n(0)$ is not proportional to $N$, which is another evidence that there are no BEC in TG gas in periodic potential.

\begin{figure}[!h]
\center
\includegraphics[scale=0.25]{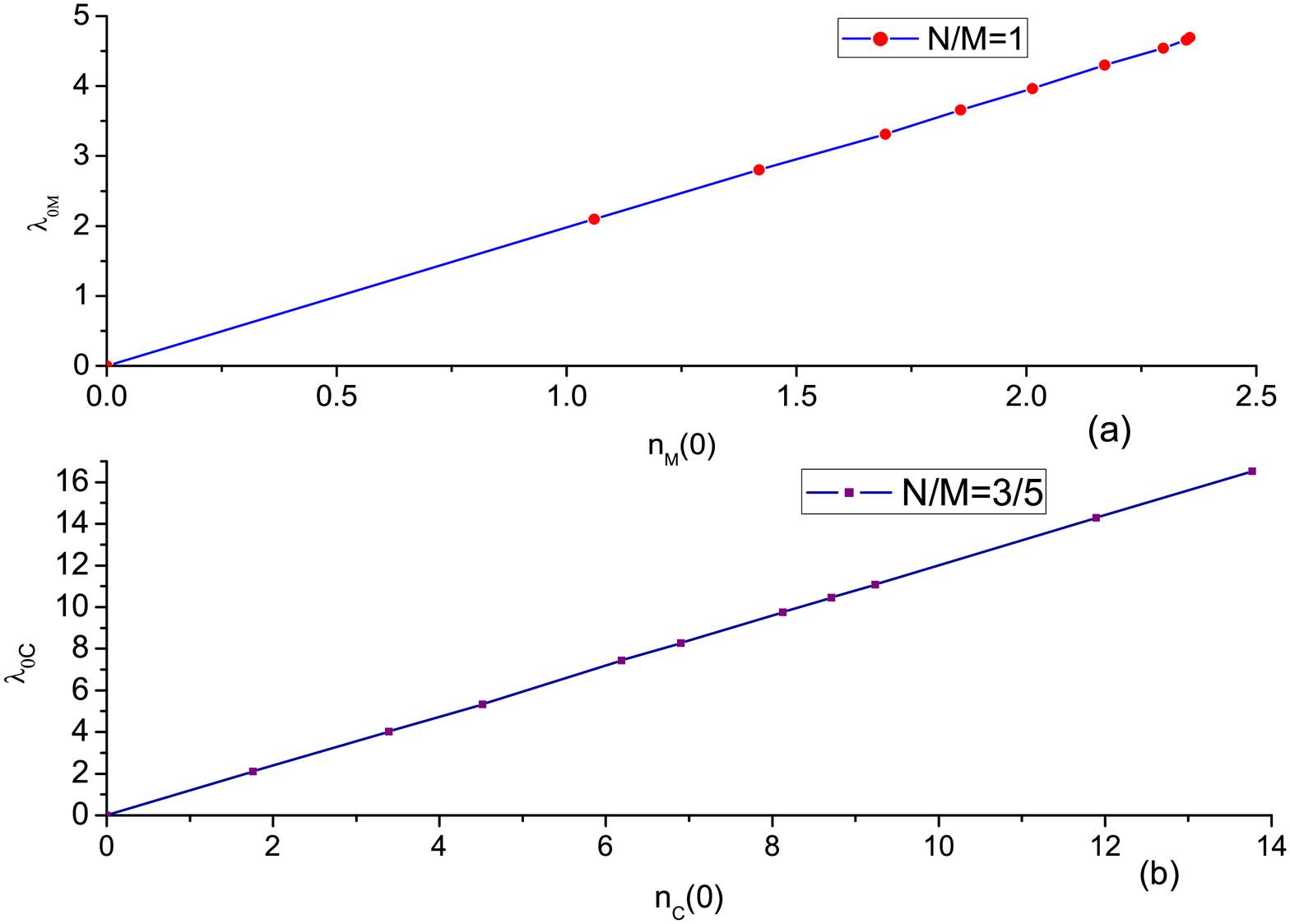}
\caption{(Color online) $\lambda_0$ as a function of $n(0)$, in commensurate (a) and incommensurate (b) cases. $q=0.1,\;\omega=1$.}
\label{fig10}
\end{figure}

The relations between the ONLNO $\lambda_0$ and $n(0)$ can be obtained by divide equation (\ref{eqn16a}) and (\ref{eqn19a}),
\begin{equation}\label{eqn21}
\frac{d\lambda_0}{dn(0)}=C,
\end{equation}
so
\begin{equation}\label{eqn22}
\lambda_0=Cn(0),
\end{equation}
where we have used the fact that $\lambda_0=n(0)=0$ as $N=0$. The relations between $\lambda_0$ and $n(0)$ are plotted in figure \ref{fig10} (a) and (b) in commensurate and incommensurate case separately. Since the momentum spectrum can be calculated by just Fourier transforming the natural orbital, it seems to me that the linear  relation between $\lambda_0$ and $n(0)$ is not surprising.

\section{Discussions and Conclusions}\label{sec4}

From equations (\ref{eqn16a}) and (\ref{eqn19a}), we know that $\lambda_0$ and $n(0)$ are proportional to $N$ if $\rho(0,\infty)$ is a constant and nonzero, so BEC occurs in this case. Otherwise $\lambda_0$ and $n(0)$ are not proportional to $N$ and there is no BEC. This inference is accord with the former work\cite{Girardeau,Penrose,Yang}.

We have studied the ONLNO, MD and off-diagonal elements of RSPDM in the ground state of TG gas in an external periodic magnetic field. The three quantities are depicted as characteristics of judging whether a Bose system exhibits BEC or not. We have found the relationships among them, which are expressed as equations (\ref{eqn16}), (\ref{eqn19}) and (\ref{eqn21}). In the special case of one dimensional periodic TG system and in the thermodynamic limit, the ONLNO $\lambda_0$ and MD $n(0)$ are proportional to $N^0$ in commensurate case while to $\sqrt{N}$ in incommensurate case.

\section*{Acknowledgment}

This work is financially supported by the National Natural Science Foundation of China under Grant No. 10975125.



\bibliographystyle{model1a-num-names}
\bibliography{<your-bib-database>}



\end{document}